\documentclass[reprint,
superscriptaddress,
 amsmath,amssymb,
 aps,
prb,
]{revtex4-1}

\usepackage{graphicx}
\usepackage{dcolumn}
\usepackage{bm}
\usepackage{color}

\begin{document}
\title{Structural and magnetic properties of the quasicrystal approximant Au$_{65}$Ga$_{21}$Tb$_{14}$ }
\author{Kazuhiro Nawa}
\affiliation{Institute of Multidisciplinary Research for Advanced Materials, Tohoku University,
  2-1-1 Katahira, Aoba, Sendai 980-8577 Japan}

\author{Maxim Avdeev}
\affiliation{Australian Centre for Neutron Scattering, Australian Nuclear Science and Technology Organisation,
  New Illawarra Rd, Lucas Heights, NSW 2234, Australia}
\affiliation{School of Chemistry, The University of Sydney, Sydney, NSW 2006, Australia}

\author{Asuka Ishikawa}
\affiliation{Research Institute for Science and Technology, Tokyo University of Science,
Tokyo 125-8585, Japan}

\author{Hiroyuki Takakura}
\affiliation{Faculty of Engineering, Hokkaido University, Sapporo, Hokkaido 060-8628, Japan}

\author{Chin-Wei Wang}
\affiliation{National Synchrotron Radiation Research Center, Hsinchu 30076, Taiwan}
\affiliation{Australian Centre for Neutron Scattering, Australian Nuclear Science and Technology Organisation,
  New Illawarra Rd, Lucas Heights, NSW 2234, Australia}

\author{Daisuke Okuyama}
\affiliation{Institute of Multidisciplinary Research for Advanced Materials, Tohoku University,
  2-1-1 Katahira, Aoba, Sendai 980-8577 Japan}

\author{Ryo Murasaki}
\affiliation{Institute of Multidisciplinary Research for Advanced Materials, Tohoku University,
  2-1-1 Katahira, Aoba, Sendai 980-8577 Japan}

\author{Ryuji Tamura}
\affiliation{Department of Materials Science and Technology, Tokyo University of Science,
 Tokyo 125-8585, Japan}

\author{Taku J Sato}
\affiliation{Institute of Multidisciplinary Research for Advanced Materials, Tohoku University,
  2-1-1 Katahira, Aoba, Sendai 980-8577 Japan}

\begin{abstract}
Magnetic properties of the quasicrystal approximant Au$_{65}$Ga$_{21}$Tb$_{14}$ were investigated by the magnetization and the neutron diffraction experiments.
The temperature dependences of the magnetic susceptibility and the magnetization curve indicate dominant ferromagnetic interactions, while the whirling antiferromagnetic order was found from the neutron diffraction experiments.
In the antiferromagnetic phase, the magnetic moments are aligned almost perpendicular to a pseudo five-fold symmetry axis, which should correspond to the easy-axis direction of a Tb atom.
The magnetic properties similar to those of Au$_{72}$Al$_{14}$Tb$_{14}$ in spite of the difference in the electron-per-atom ratio suggest the robustness of the easy-axis anisotropy against the chemical substitution of the nonmagnetic atoms.
\end{abstract}

\maketitle
\section{Introduction}
Magnetism in quasicrystals and quasicrystal approximants has attracted interest due to the expectation of nontrivial ground states.
A study in the magnetic quasicrystal and quasicrystal approximants has been accelerated since the stable binary quasicrystals and quasicrystal approximants were discovered\cite{Tsai, Goldman}.
The occurrence of the long-range antiferromagnetic order was observed in the 1/1 quasicrystal approximant Cd$_6R$ ($R$ : rare-earth elements) \cite{CdTb} and confirmed by the resonant X-ray scattering\cite{CdTb2} and neutron diffraction experiments\cite{CdTb3}. 
However, the reliable magnetic structure has not been obtained so far due to the presence of the structural phase transition:
orientational ordering of a Cd$_4$ tetrahedron, which locates at the center of each Tsai-type cluster, lowers the crystal symmetry and induces several structural domains\cite{orderdisorder, orderdisorder2}.
It was thus desired to investigate quasicrystal approximants without the structural phase transition for better understanding of their magnetic properties.

Recent neutron diffraction experiments have revealed that a few magnetic quasicrystal approximants\cite{AuAlTb1, AuAlTb1_2, AuSiTb1, AuSiTb3} show the noncollinear magnetic structures\cite{AuAlTb2, AuSiTb2}.
These quasicrystal approximants belongs to the space group of $Im\overline{3}$ as other 1/1 quasicrystal approximants,
and do not exhibit structural phase transitions down to low temperatures.
Magnetic Tb$^{3+}$ ions form nearly icosahedral clusters.
In the Au-Al-Tb 1/1 quasicrystal approximant Au$_{72}$Al$_{14}$Tb$_{14}$,
magnetic moments are aligned almost tangential to the cluster surface.
They follow a threefold symmetry around the [111] axis,  leading to the whirling arrangement.
The neighboring magnetic clusters are coupled antiferromagnetically\cite{AuAlTb2}.
In addition, the Au-Si-Tb 1/1 quasicrystal approximant Au$_{70}$Si$_{17}$Tb$_{13}$ exhibits the magnetic structure associated with that of Au$_{72}$Al$_{14}$Tb$_{14}$.
The former magnetic structure can be roughly derived by reversing the half of the magnetic moments of the latter magnetic structure.
As a result, the large spontaneous magnetic moments appear along the [111] direction\cite{AuSiTb2}.

The previous studies indicate that competing magnetic interactions and formation of the magnetic clusters are key ingredients to stabilize the noncollinear magnetic structure.
Accordingly, in principle, a variety of noncollinear magnetic structures could be realized by tuning the magnetic interactions or the magnetic anisotropy of the Tb$^{3+}$ ions\cite{Watanabe2, phasediagram}.
Since magnetic interactions are dominated by the RKKY mechanism, magnetic interactions within an icosahedral cluster could be adjusted by tuning an electron-per-atom ratio (e/a)\cite{AuAlTb1, AuAlTb1_2, approximants}.
In addition, changing a local environment around Tb$^{3+}$ ions could result in the change in the easy-anisotropy axis\cite{Watanabe1, Watanabe2}.
Magnetic properties of quasicrystal approximants with different compositions should reflect the change in the e/a and the easy-anisotropy axis.
It is necessary to reveal magnetic structures of quasicrystal approximants to understand the general trend in the magnetic quasicrystal approximants and quasicrystals.

In this study, we report the magnetic properties and magnetic structure of the Au-Ga-Tb 1/1 quasicrystal approximant\cite{AuGaTb}.
We focus on the magnetic properties of Au$_{65}$Ga$_{21}$Tb$_{14}$, which belongs to the space group of $Im\overline{3}$ as others and exhibits e/a of 1.70 that locates near
the boundary of the antiferromagnetic and ferromagnetic phases.
This e/a is much larger than that of 1.56 for Au$_{72}$Al$_{14}$Tb$_{14}$, which is located at near the other boundary of the antiferromagnetic phase\cite{approximants}.
In addition, its composition is very close to the magnetic quasicrystal that exhibits a long-range magnetic order\cite{AuGaTbquasicrystal}. 
The powder neutron diffraction experiments revealed the whirling antiferromagnetic arrangement of the magnetic moments in Au$_{65}$Ga$_{21}$Tb$_{14}$,
which is similar to that observed in Au$_{72}$Al$_{14}$Tb$_{14}$\cite{AuAlTb2}.
The whirling antiferromagnetic order realized in the wide range of e/a indicates its robustness against the chemical substitution. 

\begin{figure}[t]
\includegraphics[width=8cm]{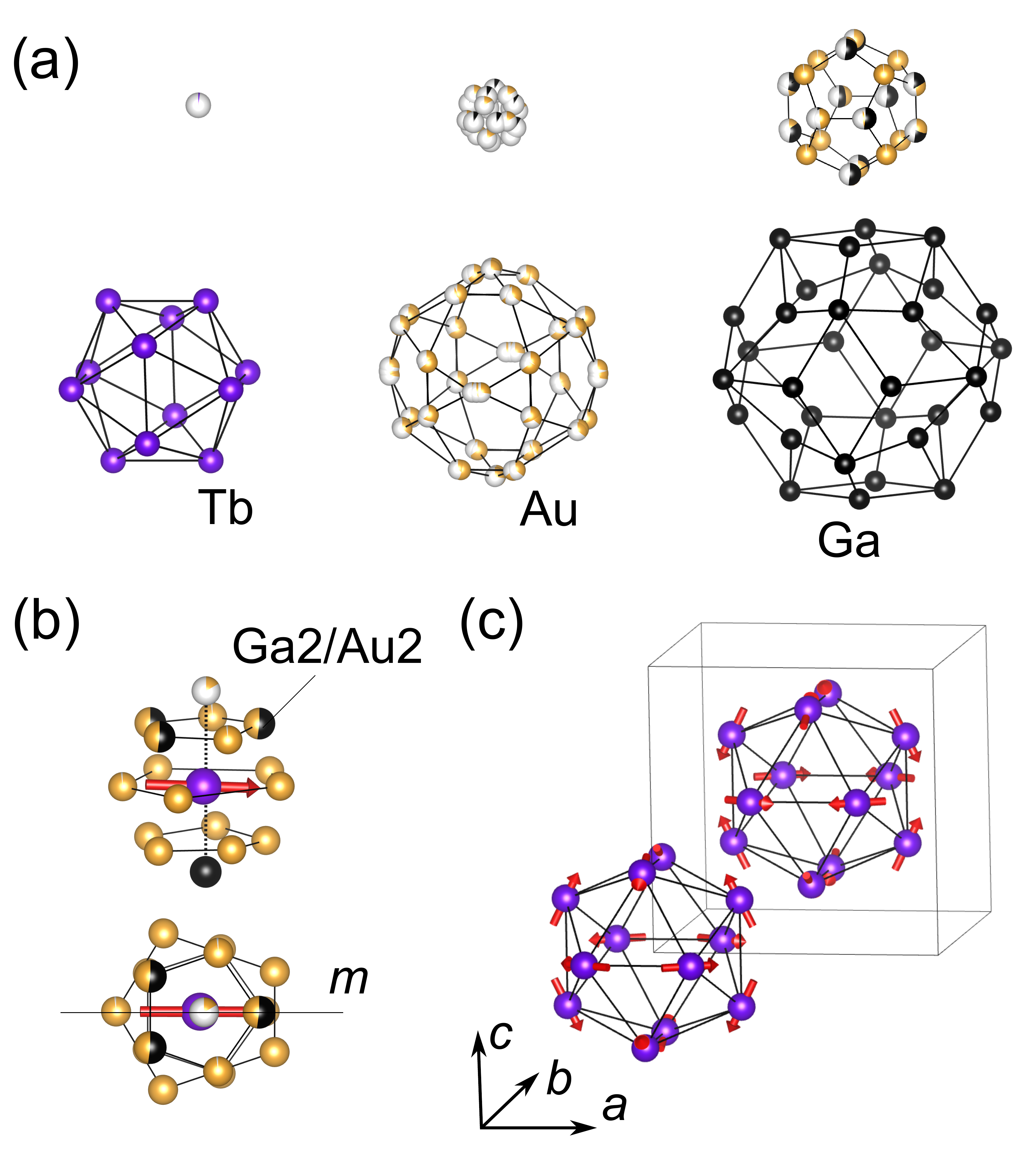}
\caption{\label{cluster} (a) Successive shells of a Tsai-type cluster in Au$_{65}$Ga$_{21}$Tb$_{14}$ obtained from single crystal X-ray diffraction.
(b) Local environment around a Tb atom.
The dashed and solid lines passing through the Tb atom represent the pseudo five-fold rotation axis and the mirror plane, respectively. 
The red arrow indicates the direction of the magnetic moment in the ordered phase.
A few inequivalent sites induced by disorder are merged into a single site for clarity.
(c) Antiferromagnetic order in Au$_{65}$Ga$_{21}$Tb$_{14}$. The whirling arrangement is formed around [111] direction.
Figures were made by using VESTA software\cite{VESTA}.}
\end{figure}

\begin{figure}[h]
\includegraphics[width=6cm]{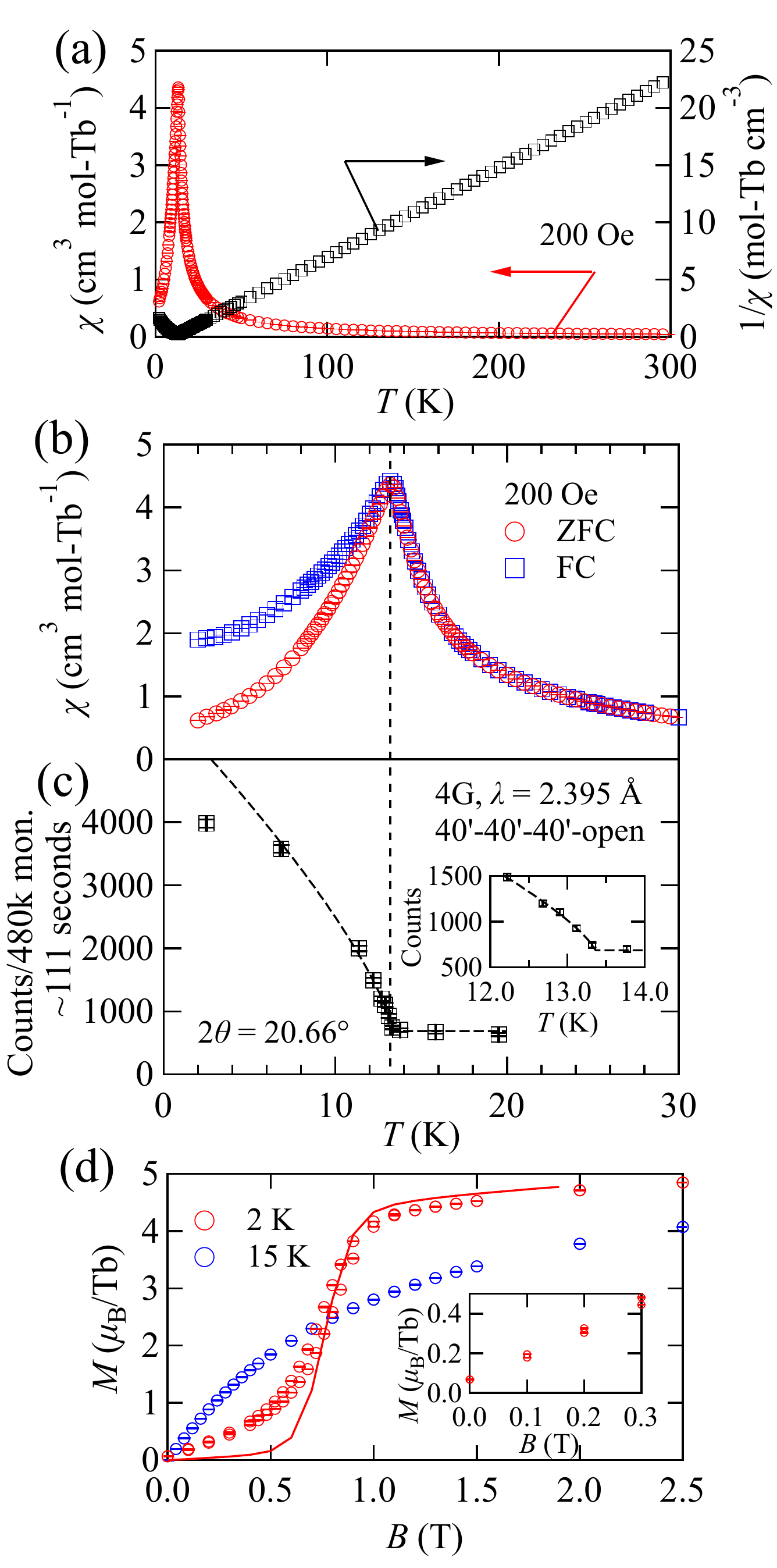}
\caption{\label{chiT} (a) Temperature dependence of the magnetic susceptibility and the inverse susceptibility. 
(b) Magnetic susceptibility below 30~K measured by zero field cooling and field cooling protocols. 
(c) Temperature dependence of the 210 magnetic reflection intensity at 20.66$^\circ$. The dashed curve represents a fit to the power law function (see text for details).
The inset shows the temperature dependence focused on near $T_\mathrm{N}$.
(d) Magnetization curve at 2 and 15~K. The red solid curve represents the simulated curve of a single cluster with the ferromagnetic interactions $J_1$ = 0.02~K and $J_2$ = 0.13~K. The inset shows the enlarged view near zero field.}
\end{figure}

\section{Experiments}
Polycrystalline samples of Au$_{65}$Ga$_{21}$Tb$_{14}$ were synthesized by arc melting using high purity elements of Au, Ga, and Tb as starting materials.
Then the alloy was annealed at 1073~K for 50~h under an Ar atmosphere, followed by quenching into chilled water.
The sample quality was confirmed by powder X-ray diffraction experiments with Cu $K_{\alpha}$ radiation (Rigaku MiniFlex600 and Rigaku Ultima IV).
The magnetic properties were investigated using  a SQUID magnetometer (Quantum Design MPMS).

The crystal structure analysis was carried out for a single crystal picked from polycrystalline sample with an XtaLAB Synergy-R single-crystal diffractometer equipped with Hybrid Pixel Array Detector (HyPix6000, Rigaku) using Mo $K_\alpha$ radiation ($\lambda$ = 0.71073 \AA).
Indexing and integration of diffraction intensities were performed utilizing CrysAlisPRO software\cite{CAPRO}.
Initial structure model was obtained by a dual-space method using SHELXT\cite{SHELXT}.
Subsequent structure refinements were conducted by using SHELXL\cite{SHELXL}.

To investigate the magnetic structure, the neutron powder diffraction experiments were performed using the high-resolution powder diffractometer ECHIDNA installed at the OPAL reactor, Australian Nuclear Science and Technology Organisation~\cite{ECHIDNA}.
Neutrons with $\lambda = 2.4395$~\AA\ was selected using the Ge 331 reflections. 2~grams of the polycrystalline sample were loaded in a vanadium can with the diameter of 6~mm.
The bottom part of the Vanadium can was filled with Cd absorber to adjust the sample position.
The sample can was set on a cold head of a closed cycle $^4$He refrigerator with the base temperature, 4 K. The intensities were collected at the base temperature of 4~K and 20~K. 
In addition, temperature dependence of a magnetic reflection was investigated using the thermal-neutron triple-axis spectrometer 4G-GPTAS installed at the JRR-3 reactor, Tokai, Japan.
The spectrometer was operated in the double-axis mode without using the analyzer.
The wavelength was set to $\lambda = 2.359$~\AA\ using pyrolytic graphite (PG) 002 reflections. The horizontal collimations of 40'-40'-40' were employed.
2 grams of the powder sample was sealed in the Al sample can with the He exchange gas, and was cooled down to 2.5 K using a closed cycle $^4$He refrigerator.

\begin{table}[t]
 \caption{\label{structure} Crystallographic data and refinement parameters of the single crystal X-ray structure analysis.}
 \label{refine}
\begin{ruledtabular}
\begin{tabular}{lc}
Space group & $Im\overline{3}$ \\
$a$ (\AA)  & 14.72530(10) \\ 
$V$ (\AA$^3$)  & 3192.95(7) \\ 
Calculated density (g/cm$^3$)  & 8.577\\ 
Temperature (K) & 293 \\
$h$ range  & $-20 \leq h \leq 19$\\ 
$k$ range  & $-20 \leq k \leq 20$\\ 
$l$ range  & $-20 \leq l \leq 19$ \\ 
$F$(000) & 6696 \\ 
Reflections collected & 48241 \\ 
Independent reflections & 897\\ 
Data\footnotemark[1]/Restraints/parameters & 869/0/70\\ 
$R$ indices for observed reflections ($R$, $R_\mathrm{w}$)\footnotemark[2] & 0.0208, 0.0370\\ 
Goodness of fit on $F$  & 1.336\\ 
\end{tabular}
\end{ruledtabular}
\footnotetext[1]{Reflections with $I > 2\sigma$ are used for the refinement}
\footnotetext[2]{$R = \sum ||F_\mathrm{obs}|^2 - |F_\mathrm{calc}|^2|/ \sum |F_\mathrm{obs}|^2,
\mathrm{w}R = \sum w (|F_\mathrm{obs}|^2 - |F_\mathrm{calc}|^2)/ \sum w |F_\mathrm{obs}|^2,
w = 1/\{ \sigma^2 (|F_\mathrm{obs}|^2) + 247.1726 (|F_\mathrm{obs}|^2 + 2|F_\mathrm{calc}|^2)/3 \}$ }
\end{table}

\section{Results and Discussions}
The single crystal X-ray diffraction experiments have revealed some differences in the crystal structure of Au$_{65}$Ga$_{21}$Tb$_{14}$ compared with that of Au$_{72}$Al$_{14}$Tb$_{14}$.
Successive clusters formed by Au, Ga, and Tb atoms are illustrated in Fig.~\ref{cluster}(a) and
the structural parameters are summarized in Table~\ref{atom}.
First, the occupancy of the Ga atom is substantially larger at a Ga2 site, which is one of the nearest-neighbor nonmagnetic site of the Tb atom. 
The atomic position of the Ga2 site is illustrated in Fig.~\ref{cluster}(b).
The occupancy is found as 0.513(7) in Au$_{65}$Ga$_{21}$Tb$_{14}$, while it is decreased down to 0.098(4) for the corresponding Al2 site (including both 2A and 2B) in Au$_{72}$Al$_{14}$Tb$_{14}$\cite{AuAlTb1_2}.
Second, the lattice constant of 14.7361~\AA \ for Au$_{65}$Ga$_{21}$Tb$_{14}$ (from the powder XRD experiments, see the supplementary information)
is slightly smaller than 14.7753~\AA \ for Au$_{72}$Al$_{14}$Tb$_{14}$\cite{AuAlTb1_2}.
The decrease of the lattice constant can be attributed to the decrease in the occupancy of Au atoms, since the atomic size of Au is larger than those of Ga and Al.
On the other hand, the distances of Tb--Tb atoms within the cluster are estimated as 5.5043(12) and 5.5467(7)~\AA.
This is slightly larger than that of 5.4693 and 5.5297~\AA \ in Au$_{72}$Al$_{14}$Tb$_{14}$.
A cluster radius, which is defined by the distance of the Tb atoms from the cluster center, is 5.2671(6)~\AA \ for Au$_{65}$Ga$_{21}$Tb$_{14}$,
while that becomes 5.2475~\AA \ for Au$_{72}$Al$_{14}$Tb$_{14}$.
The nearly icosahedral cluster may be expanded by the Au atoms occupying the Au7A sites, which is near the center of the successive clusters. 

\begin{figure}[t]
\includegraphics[width=8cm]{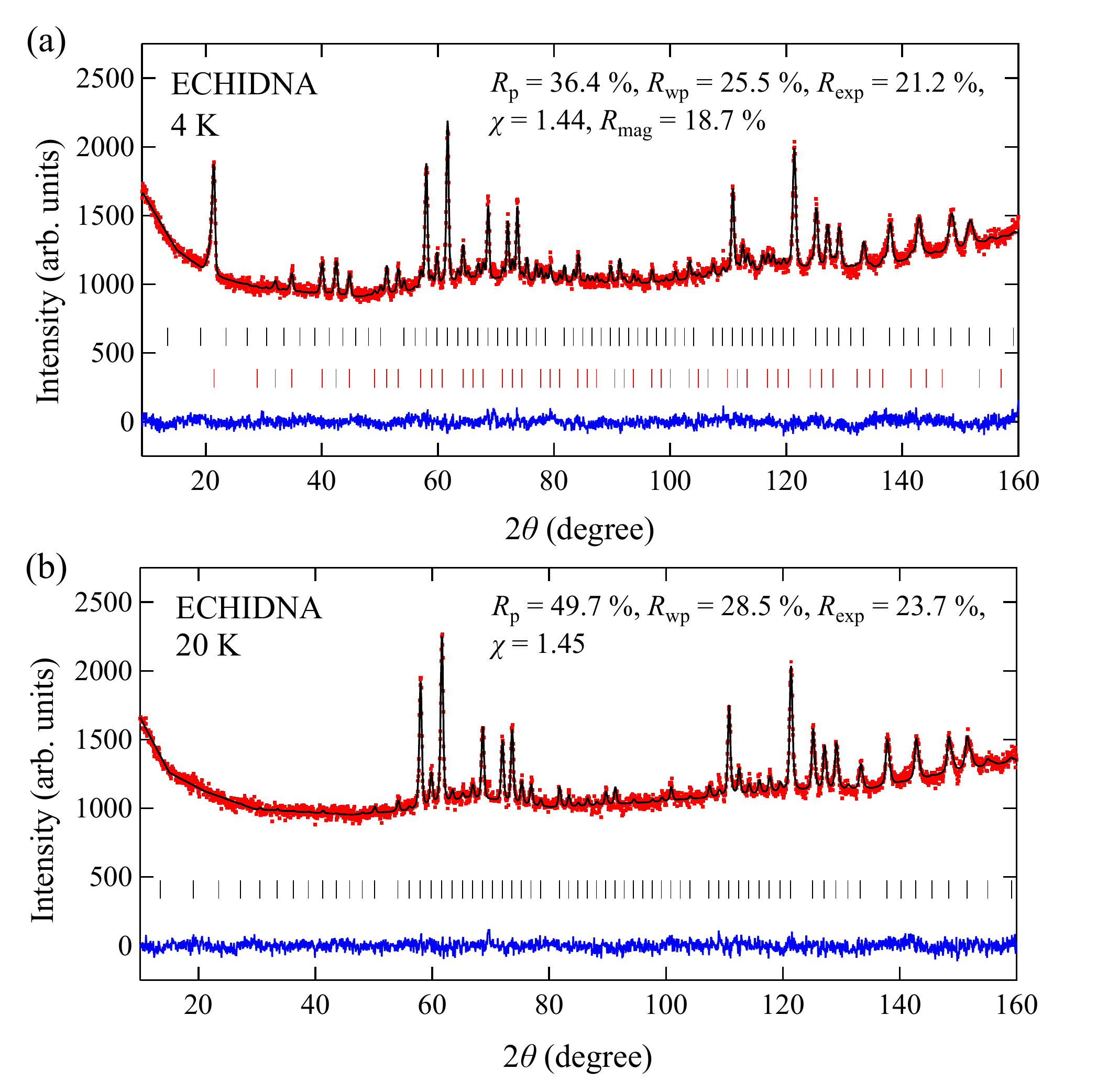}
\caption{\label{NPD}Powder neutron diffraction patterns measured at (a) 4 and (b) 20~K. Observed intensities, calculated intensities, and their difference are represented by red dots, black and blue curves, respectively.
The position of the nuclear and magnetic reflections are indicated by black and red solid lines.}
\end{figure}

\begin{table*}[t]
\caption{\label{Structure}Structure parameters of Au$_{65}$Ga$_{21}$Tb$_{14}$ determined from the single-crystal X-ray structure analysis.
The atomic coordinates are represented by fractional coordinates.
The isotropic displacement parameters were adopted for atoms with a small occupancy.
The equivalent isotropic (isotropic) displacement parameters $U_\mathrm{eq}$ ($U_\mathrm{iso}$) are listed in a unit of \AA $^2$.}
\label{atom}
\begin{center}
\begin{tabular}{lccccccccccc}
\hline
atom & site & $x$ & $y$ & $z$ & occupancy & 100 $U_\mathrm{eq}$ (100 $U_\mathrm{iso}^*$) \\ \hline 
Au1A & 48$h$ & 0.1115(7) & 0.3401(4) & 0.2020(4) & 0.36(5) & 0.70(7)$^*$ \\ 
Au1B & 48$h$ & 0.1014(5) & 0.3430(2) & 0.1977(4) & 0.64(5) & 0.87(4) \\ 
Ga2 & 24$g$ & 0 & 0.2345(9) & 0.0900(9) & 0.513(7) & 4.5(6) \\ 
Au2 & 24$g$ & 0 & 0.2392(2) & 0.0812(3) & 0.487(7) & 1.06(4) \\ 
Au3 & 24$g$ & 0 & 0.40396(3) & 0.35250(3) & 1 & 0.804(11) \\ 
Au4 & 16$f$ & 0.14974(2) & 0.14974(2) & 0.14974(2) & 0.980(4) & 1.52(2) \\ 
Ga5 & 12$e$ & 0.19441(13) & 0 & 0.5 & 1 & 0.89(3) \\ 
Au6A & 12$d$ & 0.4061(3) & 0 & 0 & 0.43(6) & 0.71(18) \\ 
Au6B & 24$g$ & 0 & 0.4039(3) & 0.0171(8) & 0.27(3) & 0.89(17)$^*$ \\ 
Tb1 & 24$g$ & 0 & 0.18690(4) & 0.30498(4) & 1 & 0.655(12) \\ 
Ga8 & 8$c$ & 0.25 & 0.25 & 0.25 & 1 & 1.91(5) \\ 
Au7A & 24$g$ & 0 & 0.0674(5) & 0.0806(6) & 0.162(6) & 5.0(2)$^*$ \\ 
Ga7B & 24$g$ & 0 & 0.0891(14) & 0.0464(17) & 0.087(14) & 1.4(7)$^*$ \\ 
Tb2 & 2$a$ & 0 & 0 & 0 & 0.026(14) & 3.0(4)$^*$ \\ 
\hline
\end{tabular}
\begin{tabular}{lcccccccc}
\hline
atom & site & 100 $U_\mathrm{11}$ & 100 $U_\mathrm{22}$ & 100 $U_\mathrm{33}$ & 100 $U_\mathrm{12}$ & 100 $U_\mathrm{23}$ & 100 $U_\mathrm{31}$ & 100 $U_\mathrm{eq}$ \\ \hline 
Au1B & 48$h$ & 0.67(8) & 0.76(5) & 1.17(5) & 0.14(3) & 0.12(5) & 0.12(5) & 0.87(4) \\ 
Ga2 & 24$g$ & 4.1(7) & 6.1(8) & 3.2(6) & 1.3(4) & 0 & 0 & 4.5(6) \\ 
Au2 & 24$g$ & 0.51(8) & 1.42(8) & 1.25(9) & $-$0.19(6) & 0 & 0 & 1.06(4) \\ 
Au3 & 24$g$ & 0.81(2) & 0.70(2) & 0.90(2) & 0.123(15) & 0 & 0 & 0.804(11) \\ 
Au4 & 16$f$ & 1.52(2) & 1.52(2) & 1.52(2) & 0.049(13) & 0.049(13) & 0.049(13) & 1.52(2) \\ 
Ga5 & 12$e$ & 1.03(8) & 0.60(8) & 1.05(8) & 0 & 0 & 0 & 0.89(3) \\ 
Au6A & 12$d$ & 0.80(18) & 0.8(6) & 0.54(17) & 0 & 0 & 0 & 0.71(18) \\ 
Tb1 & 24$g$ & 0.68(2) & 0.47(2) & 0.82(2) & 0.075(18) & 0 & 0 & 0.655(12) \\ 
Ga8 & 8$c$ & 1.91(5) & 1.91(5) & 1.91(5) & $-$0.01(6) & $-$0.01(6) & $-$0.01(6) & 1.91(5) \\ 
\hline
\end{tabular}
\end{center}
\end{table*}

The temperature dependences of the magnetic susceptibility and the magnetization curve indicate the occurrence of an antiferromagnetic order in spite of dominant ferromagnetic interactions.
At high temperatures, the magnetic susceptibility (Fig.~\ref{chiT}(a)) increases with decreasing temperature, following Curie-Weiss rule.
The Curie-Weiss fit between 50 and 300~K yields the effective magnetic moment and the  erature of 10.1(1)~$\mu_\mathrm{B}$ and 10.79(2)~K, respectively. 
At low temperatures, the magnetic susceptibility exhibits a sharp decrease below 13.2~K, suggesting the occurrence of an antiferromagnetic order (Fig.~\ref{chiT}(b)).
The deviation between the zero-field and field-cooling curves should result from a spontaneous magnetic moment,
whose magnitude is estimated as 0.068 $\mu_\mathrm{B}$ from the magnetization curve at 2~K (the inset of Fig.~\ref{chiT}(d)).
Since the ground state is quite sensitive to sample composition, it is possible that the spontaneous magnetic moment was induced from a very small portion with different composition in the sample.
At 2 K, the magnetization curve exhibits a sharp jump at 0.78~T, indicating the spin-reorientation transition.
A small field hysteresis suggests that the spin-reorientation transition is of first order.

These magnetic properties are quite close to those observed in Au$_{72}$Al$_{14}$Tb$_{14}$\cite{AuAlTb1, AuAlTb1_2}.
The transition temperature of 13.2~K is comparable to 11.8~K of Au$_{72}$Al$_{14}$Tb$_{14}$\cite{AuAlTb1}.
On the other hand, the Weiss temperature of 10.79~K is a few times larger than 4.2~K of Au$_{72}$Al$_{14}$Tb$_{14}$,
indicating the enhancement in the ferromagnetic interaction.
Simultaneously, the transition field of 0.78~T becomes almost half of 1.36~T of Au$_{72}$Al$_{14}$Tb$_{14}$\cite{AuAlTb1}.
This reflects that the value of e/a is close to the phase boundary of a ferromagnetic ground state.

Occurrence of the antiferromagnetic order is clearly noticed in comparison with the neutron diffraction patterns at 4~K and 20~K, as shown in Fig.~\ref{NPD}.
A structure refinement by the Rietveld method using the data at 20 K showed a good agreement, yielding the structure parameters consistent with
those estimated from the single-crystal X-ray diffraction.
At 4~K, magnetic reflections were observed between the nuclear reflections.
The magnetic reflections were able to be indexed with the reflection condition: $h$+$k$+$l$ = odd, where $h$, $k$, $l$ are the Miller indices.
Temperature dependence of the 210 magnetic reflection intensity is shown in Fig.~\ref{chiT}(d).
The intensity starts to increase below 13.2~K, where the magnetic susceptibility shows a sharp drop.
This supports that the body-centered cubic symmetry is broken by the occurrence of the antiferromagnetic order.
The intensity collected in the range of 12.2 and 15.8 K were fitted to a power law function proportional to
$(T_\mathrm{N} - T)^{2\beta}$ with a constant background.
The fit yields the transition temperature of $T_\mathrm{N}$ = 13.36(4)~K, which is in good agreement with 13.2~K estimated from the magnetic susceptibility.
In spite of the limited number of the data points, the critical exponent $\beta$ is also roughly estimated as 0.379(10). This value does not contradict to that expected for a 3D order\cite{beta}.

The peak positions and the intensities of the magnetic reflections are similar to those observed in Au$_{72}$Al$_{14}$Tb$_{14}$, suggesting the formation of a whirling antiferromagnetic order alike in Au$_{72}$Al$_{14}$Tb$_{14}$.
To confirm this expectation, we have analyzed the magnetic structure through the Rietveld refinement.
Candidates for initial magnetic structures are obtained using magnetic representation theory\cite{RT}.
Magnetic representations for the Tb moments are decomposed using the irreducible representations (IR) of the k-group with $k$ = (1,1,1).
The result of the decomposition and corresponding magnetic basis vectors (BVs) for all the IRs are obtained\cite{BR}. 
The good agreement is achieved by the fit based on IR2, as shown by black solid curves in Fig.~\ref{NPD}(a).
Figure~\ref{cluster}(c) illustrates the magnetic order obtained from the refinement.
The magnetic moments follow the threefold rotation symmetry around the [111] direction, forming the whirling magnetic order as that observed in Au$_{72}$Al$_{14}$Tb$_{14}$\cite{AuAlTb2}.
Each magnetic moment is required to remain in a local mirror plane by symmetry.
In addition, it is aligned almost perpendicular to a pseudo five-fold symmetry axis, as shown in Fig.~\ref{cluster}(b).
The whirling magnetic order is represented by the two coefficients of the basis vectors.
They are estimated to be 7.04(12) and $-$3.49(17) $\mu_\mathrm{B}$ in Au$_{65}$Ga$_{21}$Tb$_{14}$.
The moment direction is almost the same as that of Au$_{72}$Al$_{14}$Tb$_{14}$\cite{AuAlTb2}.
The fit yields the magnitude of the magnetic moment as 7.86(13) $\mu_\mathrm{B}$.

\section{Discussions}
The direction of the magnetic moment should reflect the easy-axis anisotropy of the Tb atom induced by the crystalline electric field.
Indeed, the easy-axis nature for the magnetic moment on the Tb atom
has been indicated in Tb$_6$Cd and Au$_{70}$Si$_{17}$Tb$_{13}$.
Analysis on the inelastic neutron spectrum assuming a local pentagonal symmetry\cite{petagon} suggests that the easy axis direction is along the pseudo five-fold symmetry axis in Cd$_6$Tb\cite{Cd6Tb_CEF}.
On the other hand, neutron diffraction and inelastic neutron scattering study indicates that it is perpendicular to the pseudo five-fold symmetry axis in Au$_{70}$Si$_{17}$Tb$_{13}$\cite{AuSiTb2}.
As shown in Fig.~\ref{cluster}(b), the direction of the magnetic moment observed in Au$_{65}$Ga$_{21}$Tb$_{14}$ (which is almost equivalent to that in Au$_{72}$Al$_{14}$Tb$_{14}$) is perpendicular to the pseudo five-fold symmetry axis.
Thus, the easy-axis direction should be also perpendicular to the pseudo five-fold symmetry axis in these compounds.
According to the crystalline electric field calculations based on the point charge model, the easy-axis direction can range between the parallel and the perpendicular direction to the pseudo five-fold symmetry axis, depending on the ratio of the effective valences of ligand atoms\cite{Watanabe2}.
The similarity in the direction of magnetic moments between Au$_{72}$Al$_{14}$Tb$_{14}$ and Au$_{65}$Ga$_{21}$Tb$_{14}$ indicates
that the ground state wave function of Tb$^{3+}$ ions is almost unchanged in spite of large difference in the occupancy of the Ga2 (Au2) atom.
The change in the crystalline electric field may be suppressed by the screening effect of conduction electrons.

The macroscopic magnetic properties suggest that ferromagnetic interactions are enhanced in Au$_{65}$Ga$_{21}$Tb$_{14}$ compared with those of Au$_{72}$Al$_{14}$Tb$_{14}$.
The possible nearest neighbor ($J_1$) and next-nearest neighbor interactions ($J_2$) were estimated by reproducing the magnetization curve. 
The simulation was performed based on a single cluster composed by Ising spins, following the procedure presented in ref.~\onlinecite{AuAlTb2}.
The cluster is regarded as an icosahedra with two types of Heisenberg interactions,
\begin{equation}
H = -J_1 \sum_{\mathrm{NN}} \mathbf{S}_i \cdot \mathbf{S}_j - J_2 \sum_{\mathrm{NNN}} \mathbf{S}_i \cdot \mathbf{S}_j + g\mu_\mathrm{B} H \sum_i \mathbf{S}_i.
\end{equation}
where $J_1$ and $J_2$ represent magnetic interactions coupled to five nearest and five next-nearest neighbor sites, respectively.
The ferromagnetic interactions of $J_1$ =~0.0221~K and $J_2$ =~0.132~K well reproduce the experimental data, as shown in the red solid curve in Fig.~\ref{chiT}(d).
In addition, they yield the Weiss temperature of $\theta_\mathrm{W} = 5J(J+1)(J_1+ J_2)/3$ = 10.8~K ($J$ =~6), consistent with that estimated experimentally.
Furthermore, they are compatible with the occurrence of the antiferromagnetic order, which is stable under the condition of $J_1 < J_2/2$\cite{approximants, AuAlTb2}.

Let us compare magnetic properties of Au$_{65}$Ga$_{21}$Tb$_{14}$ and Au$_{72}$Al$_{14}$Tb$_{14}$, both of which are confirmed to have the similar magnetic structure.
The e/a of the two compounds are 1.70 and 1.56, respectively.
The large e/a is qualitatively consistent with the enhanced ferromagnetic interactions,
according to the phase diagram established from the magnetization measurements on the series of magnetic quasicrystal approximants\cite{AuAlTb1, AuAlTb1_2, approximants}.
The further increase in the e/a would enhance ferromagnetic interactions, and result in the ferromagnetic order as that observed in Au$_{70}$Si$_{17}$Tb$_{13}$\cite{AuSiTb2}.
The discovery of the whirling magnetic order in Au$_{65}$Ga$_{21}$Tb$_{14}$, which e/a locates near the boundary of the antiferromagnetic and ferromagnetic phases,
strongly supports the description of nearly icosahedral clusters composed by magnetic moments with strong easy-axis anisotropy and competing ferromagnetic interactions between them. 
On the other hand, it is remaining unclear how magnetic interactions and magnetic structures are modified by tuning the value of e/a.
It would be an interesting subject to investigate effective magnetic exchanges by inelastic neutron scattering experiments to understand this mechanism.

\section{Conclusion}
The whirling antiferromagnetic magnetic order was found in the quasicrystal approximant Au$_{65}$Ga$_{21}$Tb$_{14}$.
The magnetization measurements and neutron diffraction experiments indicate that the antiferromagnetic order is induced by the ferromagnetic next-nearest neighbor interactions and the easy-axis anisotropy of the Tb atom.
In spite of the large difference in the occupation ratio of chemically disordered atoms at some ligand sites, the direction of the magnetic moment in the antiferromagnetic phase is almost the same as that in Au$_{72}$Al$_{14}$Tb$_{14}$.
The little influence of the chemical substitution on the easy-axis direction suggests that the crystalline electric field can be screened by conduction electrons in magnetic quasicrystal approximants.
.

\section{acknowledgements}
This work was supported by Grants-in-Aid for Early-career scientists (No.~20K14395), Scientific Research on Innovative Areas (Nos.~19H05818, 19H05819 and 20H05261),
Scientific Research (B) (No.~19H01834) and Fund for the Promotion of Joint International Research (Fostering Joint International Research) (No.~18KK0150) from the Japan Society for the Promotion of Science,
and the CORE Laboratory Research Program “Dynamic Alliance for Open Innovation Bridging Human, Environment and Materials” of the Network Joint Research Center for Materials and Device”.
The experiments at JRR-3, as well as the travel expenses for the neutron scattering experiment at ANSTO, was partly supported by the General User Program for Neutron Scattering Experiments, Institute for Solid State Physics, University of Tokyo.
K. N. appreciates Satoshi Kameoka for his help to perform powder XRD experiments.

\bibliography{draftv1}

\end{document}